\documentclass[prl,twocolumn,superscriptaddress,longbibliography]{revtex4-1}

\usepackage{tabularx}
\usepackage{float}
\usepackage{bm}
\usepackage{graphicx}
\usepackage{amssymb}
\usepackage{amsmath}
\usepackage{mathtools}
\usepackage{color}
\usepackage{mathrsfs}
\usepackage{hyperref}

\usepackage{subfiles}

\def\kk{\mathbf{k}}

\usepackage{changes}

 \colorlet{Changes@Color}{blue}

\begin{document}

\title{Vacuum anomalous Hall effect in gyrotropic cavity}

\author{I. V. Tokatly}
\affiliation{Nano-Bio
  Spectroscopy group and European Theoretical Spectroscopy Facility (ETSF), Departamento de Pol\'imeros y Materiales Avanzados: F\'isica, Qu\'imica y Tecnolog\'ia, Universidad del
  Pa\'is Vasco, Av. Tolosa 72, E-20018 San Sebasti\'an, Spain}
 \affiliation{IKERBASQUE, Basque Foundation for Science, 48009 Bilbao, Spain}
\affiliation{Donostia International Physics Center (DIPC), 20018 Donostia-San Sebasti\'{a}n, Spain}
\affiliation{ITMO University, Department of Physics and Engineering, Saint-Petersburg, Russia}

\author{D. Gulevich}
\affiliation{ITMO University, Department of Physics and Engineering, Saint-Petersburg, Russia}

\author{I. Iorsh}
\affiliation{ITMO University, Department of Physics and Engineering, Saint-Petersburg, Russia}

\begin{abstract}
We consider the ground state of an electron gas embedded in a quantum gyrotropic cavity. We show that the light-matter interaction leads to a nontrivial topology of the many-body electron-photon wave function characterized by a nonzero Berry curvature. Physically, the latter manifests as the anomalous Hall effect, appearance of equilibrium edge/surface currents and orbital magnetization induced by vacuum fluctuations. Remarkably, closed analytical expressions for the anomalous Hall conductivity and macroscopic magnetization are obtained for the interacting many-body case.
\end{abstract}

\maketitle

Recent advances in nanotechnology allowed to increase the effective light-matter coupling to the limit when the border between the condensed matter theory and quantum optics becomes completely blurred. Since the pioneering work of Hopfield in 1958~\cite{PhysRev.112.1555}, it has been anticipated that the light-matter interaction leads to the emergence of 
hybrid quasiparticles -- polaritons, which inherit properties of both photonic and matter excitations. Since then, polaritons were studied in a vast majority of systems ranging from the single atoms and molecules to superconductors~\cite{basov2021polariton} and the emergent fields of cavity (CQED) and waveguide (WQED) quantum electrodynamics~\cite{mabuchi2002cavity,RevModPhys.90.031002} explore the fundamental quantum properties of the polaritons as well as the novel quantum information processing protocols exploiting these structures.

Until recently, most of the WQED and CQED set-ups could be adequately described within the rotating wave approximation (RWA). Under RWA the light-matter coupling Hamiltonian contains only terms preserving the total number of excitations. As a consequence, the ground state of the system comprises a photonic vacuum and the material ground state and remains unaffected by the light-matter interaction. The RWA is applicable whenever the ratio of the characteristic energy of the light-matter coupling $g$ and the photonic excitation energy $\Omega$ is negligible, $g/\Omega \ll 1$, and remains absolutely adequate approximation in most of the conventional cavity systems with the photonic excitations in the optical range $\Omega\sim 1~\mathrm{eV}$. However, for the last decade, a plethora of cavity designs where the ratio $g/\Omega$ could reach and even exceed $0.1$ have been demonstrated for the optical~\cite{chikkaraddy2016single}, terahertz~\cite{anappara2009signatures} and microwave~\cite{gu2017microwave}. The pioneering experiments boosted the interest in the so-called ultrastrong coupling regime of the light-matter interaction~\cite{kockum2019ultrastrong} (USC).

In USC regime, the terms in light-matter coupling Hamiltonian which do not preserve the total number of excitations can no longer be neglected. The ground state then becomes a mixture of the matter and photonic degrees of freedom and is characterized by non-zero values of the matter and photon occupation numbers. This may lead to substantial modifications of the material ground state and opens new route to the versatile control over the material properties via the ultrastrong coupling with the cavity electromagnetic field vacuum fluctuations, which has been recently termed \textit{cavity QED materials engineering}~\cite{hubener2020engineering}. The emergent effects range from the modification of chemical reactions~\cite{herrera2016cavity,ebbesen2016hybrid,bennett2016novel,martinez2018can,Tokatly2013PRL,ruggenthaler2014quantum,schafer2018ab} to cavity-mediated superconductivity~\cite{thomas2019exploring,curtis2019cavity,sentef2018cavity,schlawin2019cavity} and other cavity mediated phase transitions~\cite{rokaj2019quantum,ashida2020quantum,PhysRevLett.125.257604,mivehvar2021cavity,li2020manipulating}.
\begin{figure}[!h]
\includegraphics[width=0.95\columnwidth]{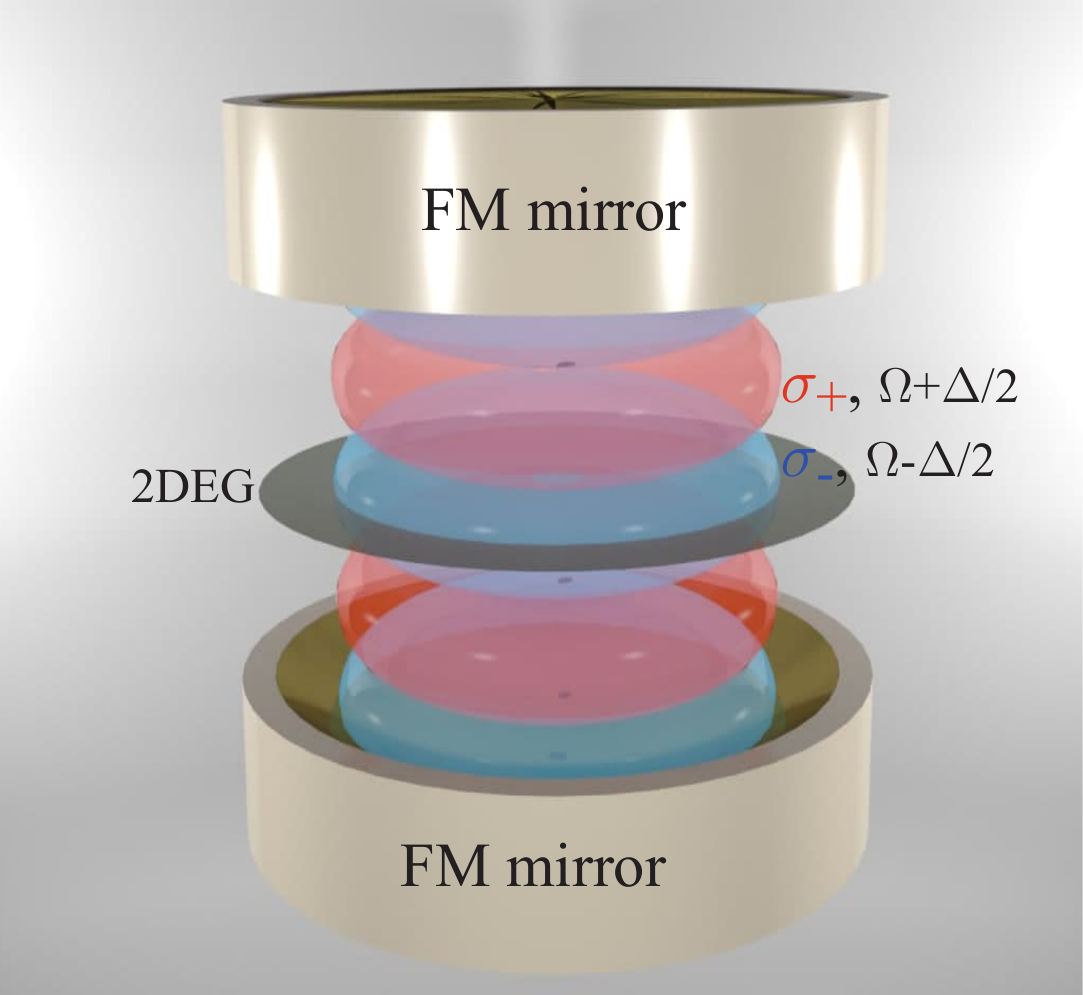}
\caption{\label{fig:f1} 
Geometry of the structure. Two dimensional electron gas is placed inside a Fabry-Perot cavity with ferromagnetic mirrors. The magnetization of the mirrors splits the energies of the circularly polarized cavity modes, which splitting induces the anomalous magnetization of 2DEG.}		
\end{figure}
The case of the spatially uniform vacuum field is particularly attractive, since it allows for the exact analytical solutions for a wide class of problems. On the other hand, some cavity mediated transitions, e.~g., superradiance, are forbidden for the spatially uniform field~\cite{PhysRevB.100.121109} (while allowed for spatially varying cavity modes~\cite{PhysRevB.102.125137}). Moreover, there is still no definite answer, whether cavity-mediated corrections to the ground state of a material are  extensive quantities and thus can affect macroscopic observables. Recently~\cite{pilar2020thermodynamics} it has been shown that this is not the case  for a collection of $N$ two-level atoms inside a single mode cavity: the cavity-mediated corrections to  macroscopic observables depend only on the effective coupling of a single two-level system to the cavity and thus vanish in the thermodynamic limit.

Here, we study an electron gas confined inside a gyrotropic (or chiral) cavity. The gyrotropy results in the energy splitting of the right- and left-circularly polarized modes. We show that vacuum fluctuations of electro-magnetic field in such cavity induce the ground state orbital magnetization of the electron gas, which is an extensive quantity, and derive the corresponding anomalous Hall conductivity of the system.

While our results are quite general, for definiteness, we consider a two-dimensional electron gas (2DEG) inside a Fabry-Perot cavity of width $D$ as shown in Fig.1. The interior of the cavity is vacuum, $\varepsilon_0=1$, and the mirrors are modelled by half spaces of a ferromagnetic metal characterized the effective permittivity tensor $\hat{\varepsilon}$
\begin{align} \label{epsilon}
    \hat{\varepsilon} = \begin{pmatrix} \varepsilon & 4i\pi\sigma_H/\omega & 0 \\ -4i\pi\sigma_H/\omega & \varepsilon & 0 \\ 0 & 0 & \varepsilon \end{pmatrix},
\end{align}
where $\varepsilon<0$ is the diagonal permittivity. The off-diagonal conductivity $\sigma_H$ is proportional to magnetization and responsible for the gyrotropy. We omit the frequency dispersion of $\varepsilon$ and $\sigma_H$ as well as material losses inside the cavity mirrors.  The equation for photon eigenmodes in the cavity reads:
\begin{align}
    \tan(\omega D/2c)=\kappa_{\pm}, \label{eq:class_Disp}
\end{align}
where $\kappa_{\pm}=\sqrt{|\varepsilon|\pm \varepsilon_H}$ and $\varepsilon_H=4\pi\sigma_H/\omega$. The modes corresponding to plus and minus sign are right and left circularly polarized modes, respectively. Equation~\eqref{eq:class_Disp} has an infinite number of solutions which correspond to the eigenfrequencies in the cavity. In the physically relevant case of $|\varepsilon|\gg 1$, the lowest eigenfrequencies are given by \textcolor{black}{$\omega_{\pm}=\Omega_0\pm 4\sigma_H/|\varepsilon|^{3/2}$, where $\Omega_0=\frac{c}{D}(\pi-2/\sqrt{|\varepsilon|})$}. 

The permittivity of Eq.\eqref{epsilon} translates to the following expression for the energy of electro-magnetic field in the gyrotropic cavity $E_{EM}= E_{EM}^{(0)}+\frac{1}{2c^2}\int d^3\mathbf{r} \sigma_H(z)[\dot{\mathbf{A}}\times \mathbf{A}]$, where $E_{EM}^{(0)}$ corresponds to the cavity with $\sigma_H=0$. We then quantize the system in the basis of eigenstates related to $E_{EM}^{(0)}$, and truncate the basis to the two lowest energy states, which are degenerate and characterized by orthogonal linear polarizations. The $x$- and $y$-components of the vector potential operator then read, 
\begin{align}
    \hat{A}_{x,y}(z)=\sqrt{\frac{\hbar c^2}{2V\Omega_0}} (\hat{a}_{x,y}+\hat{a}^{\dagger}_{x,y})\phi(z)=\sqrt{\frac{\hbar c^2}{V\Omega_0}}\hat{q}_{x,y}\phi(z),
    \label{A-def}
\end{align}
where $V=SD$ is the mode volume, $S$ is the cavity area, and $\phi(z)$ is the normalized mode profile. Operators $\hat{a}_{x,y}$ are the conventional bosonic annihilation operators. By noticing that the time derivative of the vector potential $\dot{A}_{x,y}$ is proportional to the canonical momentum $\hat{\pi}_{x,y}$, which satisfies $[\hat{q}_i,\hat{\pi}_j]=i\delta_{ij}$, we obtain the following Hamiltonian of electromagnetic field in the gyrotropic cavity, 
\begin{align} \nonumber
    H_\textrm{EM} &=\hbar\Omega_0(\hat{a}_x^{\dagger}\hat{a}_x+\hat{a}_y^{\dagger}\hat{a}_y+1)+i\hbar\Delta (\hat{a}_x^{\dagger}\hat{a}_y-\hat{a}_y^{\dagger}a_x)\\
    & = \frac{\hbar\Omega_0}{2}\hat{\bm\pi}^2 + \hbar\Delta\hat{\bm\pi}({\bm e}_z\times{\bf q}) + \frac{\hbar\Omega_0}{2}{\bf q}^2
    \label{H_EM}
\end{align}
where in the limit of $|\varepsilon|\gg 1$ the  \textcolor{black}{ gyration parameter $\Delta =4\sigma_H/|\varepsilon|^{3/2}$.} This Hamiltonian yields two circularly polarized modes with energies reproducing the eigenfrequencies of the classical problem. Equation~\eqref{H_EM} shows that in a gyrotropic cavity the photons are mapped to the excitations of a harmonic oscillator rotating at the frequency $\Delta$ or, equivalently, of an oscillator subjected to an effective magnetic field $B_{\rm eff}=2\Delta/\Omega_0$.  

The Hamiltonian of the electron gas coupled to the cavity photons is given by 
\begin{align}
    H_e = \sum_{i=1}^{N} \Big[\frac{\big(\hat{\mathbf{p}}_i-\frac{e}{c}\hat{\mathbf{A}}\big)^2}{2m} + U(\mathbf{r}_i)+\frac{1}{2}\sum_{j\ne i} V_{\mathbf{r}_i-\mathbf{r}_j}\Big],
\end{align}
where $N$ is the number of electrons,  $U$ is the external confining potential, $V_{\mathbf{r}_i-\mathbf{r}_j}$ is the direct electron-electron interaction, and the light-matter interaction is described by a minimal coupling to $\hat{\mathbf{A}}$ of eq.~\eqref{A-def}. The total electron-photon Hamiltonian is given by the sum, $H=H_{EM}+H_e$.

In the following we concentrate on the ground state of a homogeneous 2DEG with a large area $S$, such that the density $n=N/S$ remains finite even in the limit $S,N\to\infty$. Bulk properties of such systems are customary addressed by setting $U({\bf r})=0$ and imposing periodic boundary conditions, thus making the problem formally translation invariant. Further, we assume as usual that the spatial dependence of the relevant electromagnetic mode functions can be omitted, which corresponds to the dipole approximation. After these simplifications it can be immediately noticed that the cavity field couples only to the centre of mass (COM) degree of freedom of the electron gas. Moreover, the COM and the relative motions become separable. In is convenient to perform this separation in terms of 
the scaled centre of mass coordinate $\mathbf{R}=\frac{1}{\sqrt{N}}\sum_i \mathbf{r_i}$ and its conjugate momentum, $\hat{\mathbf{P}}=\frac{1}{\sqrt{N}}\sum_{i}\hat{\mathbf{p}}_i$. The total Hamiltonian then reads, 
\begin{align}
    H=H_\textrm{rel}+H_\textrm{EM}+\frac{1}{2m}(\hat{\mathbf{P}}-g_0\sqrt{N}\hat{\mathbf{q}})^2,
\end{align}
where $g_0=\left[\hbar e^2 /(SD\Omega_0)\right]^{1/2}$, and $H_\textrm{rel}$ describes the relative motion of electrons not affected by the EM field. 
 
\begin{figure}[h!]
\includegraphics[width=0.9\columnwidth]{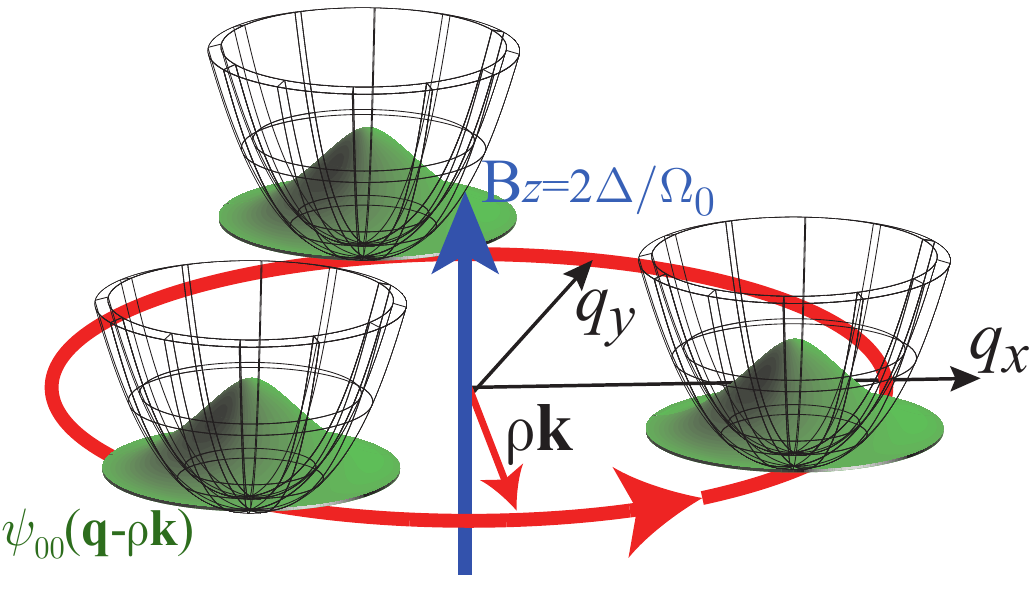}
\caption{\label{fig:Berry} 
The schematic image of the polaritonic states which are eigenfunctions of a shifted harmonic oscillator in the presence of the an effective magnetic field $B_{\rm eff}=\frac{2\Delta}{\Omega_0}$. When transported along a closed contour in $\mathbf{k}$-space they accumulate a nontrivial Berry phase.
}		
\end{figure}

Let us study a parametric dependence of the many-body eigenstates on the COM momentum. This dependence can be formally introduced via the following unitary transformation, which is equivalent to the so-called twisted boundary conditions trick \cite{XiaChaNiu2010},
\begin{align}
H_\mathbf{k}=e^{-i\mathbf{k}\mathbf{R}}He^{i\mathbf{k}\mathbf{R}}.  \label{eq:LLP}
\end{align}
Standard arguments \cite{NiuThoWu1985,XiaChaNiu2010} relate the Berry curvature $\mathcal{F}_{xy}$, associated to the ground state $|\Psi_{0,\mathbf{k}}\rangle$ of $H_\mathbf{k}$, to the anomalous Hall conductivity $\sigma_{xy}$ of 2DEG, 
\begin{align}
    \label{sigma_xy-def}
    \sigma_{xy}= \frac{e^2}{\hbar}n \mathcal{F}_{xy}|_{\mathbf{k=0}}
\end{align}
where $\mathcal{F}_{xy}=2{\rm Im} \langle\partial_{k_x}\Psi_{0,\mathbf{k}}|\partial_{k_y}\Psi_{0,\mathbf{k}}\rangle$. To proceed further we represent $H_\mathbf{k}$ of Eq.~\eqref{eq:LLP} explicitly as follows,
\begin{align}\nonumber
    &H_{\mathbf{k}}=H_\textrm{rel}+H_\textrm{EM}+\frac{1}{2m}(\hbar\mathbf{k}-g_0\sqrt{N}\mathbf{q})^2 = H_\textrm{rel}\\
    &+ \frac{\hbar\Omega_0}{2}\Big[\hat{\bm\pi} + \frac{\Delta}{\Omega_0}({\bm e}_z\times{\bf q})\Big]^2 + \frac{\hbar\Omega}{2}({\bf q}-\rho\mathbf{k})^2 + \frac{\hbar^2\mathbf{k}^2}{2m^*} \label{H_k}
\end{align}
Here $\Omega=(\tilde{\Omega}_{0}^2+\gamma^2)/\Omega_0$ with $\tilde{\Omega}_0^2=\Omega_0^2 - \Delta^2$ and $\gamma^2=\frac{g_0^2N\Omega_0}{\hbar m}$, $m^*=m(1+\gamma^2/\tilde{\Omega}_0^2)$ is the electron mass renormalized due to electron-photon interaction, and $\rho$ is given by
\begin{align}
    \rho = \frac{g_0\sqrt{N}}{m\Omega} = \sqrt{\frac{\hbar\Omega_0}{m}} \frac{\gamma}{\tilde{\Omega}_0^2+\gamma^2}.
\end{align}
From Eq.~\eqref{H_k} we see that $H_\mathbf{k}$ is nothing but the shifted Fock-Darwin Hamiltonian with a shift proportional to the COM momentum of electrons. This explains the origin of the Berry phase of the many-body polaritonic state. When $\mathbf{k}$ is moving along a contour enclosing a unit area in the $\mathbf{k}$-space, the wave function is transported in the $\mathbf{q}$-space along a contour enclosing the area $\rho^2$. The flux $\rho^2B_{\rm eff}$ of the effective magnetic field $B_{\rm eff}=\frac{2\Delta}{\Omega_0}$ gives the Berry phase accumulated in this process, see Fig.~\ref{fig:Berry}.

More formally, eigenfunctions of $H_\kk$ in Eq.~\eqref{H_k} are the products $\Psi_\kk = \Psi_{\rm rel}\Phi_\kk(\mathbf{q})$ of the wave functions $\Psi_{\rm rel}$ for the relative motion, and the polaritonic states, 
\begin{align}
\Phi_k(\mathbf{q})=e^{-i\frac{\Delta}{\Omega_0}\rho[\mathbf{e}_z\times \mathbf{k}]\mathbf{q}}\psi_{n,l}(\mathbf{q}-\rho\kk) \label{eq:Phi}
\end{align}
where $\psi_{n,l}(\mathbf{q})$ are eigenfunction of a 2D harmonic oscillator (the ground state corresponds to $n=l=0$). As the $\kk$-dependence of the wave function is known explicitly we can directly compute the Berry connection $\bm{{\cal A}}_{{\bf k}}$ as
\begin{align}
    \bm{{\cal A}}_{{\bf k}}=-i\int d\mathbf{q}\, \Phi_{k}^*(\mathbf{q})\nabla_{\mathbf{k}}\Phi_{k}(\mathbf{q}) =\frac{\Delta}{\Omega_0}\rho^2\, \mathbf{e}_z\times \mathbf{k}
\end{align}
The associated Berry curvature ${\cal F}_{xy}=(\nabla_{{\bf k}}\times\bm{{\cal A}}_{{\bf k}})_{z}$ reads,
\begin{align}
    {\cal F}_{xy}=\frac{2\Delta}{\Omega_0}\rho^2=\frac{\hbar}{m} \frac{2\Delta\gamma^2}{(\tilde{\Omega}_0^2+\gamma^2)^2},
\end{align}
as expected. We note that in the thermodynamic limit $N,S\to\infty$ at $N/S\to n$, the coupling $\gamma\to \sqrt{e^2n/mD}$ is the effective plasma frequency of electrons in the cavity. Since the Berry curvature does not depend on $\mathbf{k}$ the corresponding Hall conductivity of Eq.~\eqref{sigma_xy-def} is just
\begin{align}
    \sigma_{xy}= \frac{e^2 n}{m} \frac{2\gamma^2\Delta}{(\tilde{\Omega}_0^2+\gamma^2)^2}
\end{align}

We stress here that the electron Berry curvature emerges solely due to the electron-photon coupling. This effect is thus essentially different to the recently proposed quantum Hall effect for the graphene sheet in a magnetic cavity~\cite{Kibis1,Sentef1}. In the case of graphene, the electronic bands are initially characterized by the non-vanishing curvature, and the role of the cavity is limited to the opening of the gap in the Dirac point.

The analysis is straigtforwardly extended to the case of the infinite number of the  cavity modes spatially uniform in the 2DEG plane (the details can be found in SM). If the mode index is labelled by $\alpha$ with the corresponding $\gamma_{\alpha}$,  $\Delta_{\alpha}$, and $\Omega_{\alpha}$ the expression for the conductivity reads (see SM for details):
\begin{align}
    \sigma_{xy}= \frac{e^2 n}{m}\sum_{\alpha} \frac{2\gamma_{\alpha}^2\Delta_{\alpha}}{\tilde{\Omega}_{\alpha}^4(1+\sum_{\beta}\gamma_{\beta}^2/\tilde{\Omega}_{\beta}^2)^2}
\end{align}
For a Fabry-Perot cavity, this expression results in 
\begin{align}
     \sigma_{xy}= \frac{e^2 n}{m}\frac{\pi^4}{90} \frac{2\gamma_{1}^2\Delta_{1}}{(\tilde{\Omega}_1^2+\frac{\pi^2}{6}\gamma_{1}^2)^2},
\end{align}
where index $1$ corresponds to the fundamental Fabry-Perot mode. As it can be seen, inclusion of the multiple modes only weakly renormalizes the Hall conductivity. The Hall conductivity of 2DEG induced by the cavity modes can be rewritten as
\begin{align}
    \sigma_{xy}/c\approx { \frac{\varepsilon_H}{|\varepsilon|^{\frac{3}{2}}}}
    \frac{1}{\left(1+\frac{\tilde{\Omega}^2}{\gamma^2}\right)^2},
\end{align}
which directly relates it to the Hall permittivity of the mirrors. As can be seen, as $\tilde{\Omega}/\gamma$ vanishes, the denominator approaches a finite value, but at the same time, at $\Omega\rightarrow 0$, the Drude permittivity $|\varepsilon|\rightarrow\infty$ and thus in the dc limit, the effect vanishes. 

It is natural to expect \cite{XiaChaNiu2010} that a nontrivial Berry curvature and the bulk anomalous Hall effect should be accompanied with the orbital magnetization of the sample. Indeed, by adopting the approach of~\cite{Shi2007,Essin2010,GonZwa2011} we obtain the following ground state magnetization $M_z$
\begin{align} \label{M_z}
    M_z= -\frac{e}{mc} P_0 \frac{2\gamma^2\Delta}{(\tilde{\Omega}_0^2+\gamma^2)^2},
\end{align}
where $P_0$ is the ground state pressure of 2DEG. The rigorous derivation is presented in SM, however this result can be understood from the following simple argument. Because of the anomalous Hall effect, the gradient $\nabla U$ of the confining potential near the edges will generate the edge charge current. For example, for the boundary along $x$-axis, assuming a sufficiently smooth edge potential, and applying the Hall relation locally we get the edge current density,
\begin{align}
    j_x = -\frac{1}{e}\sigma_{xy}\partial_y U=-\frac{e}{\hbar} \mathcal{F}_{xy}\,n\partial_y U =
    \frac{e}{\hbar}\mathcal{F}_{xy}\partial_yP_0
\end{align}
where we used the force balance condition $n\nabla U= -\nabla P_0$. By integrating the above expression across the boundary (from the interior to the exterior of the sample) we get the net edge current $I_{\rm edge}= \frac{e}{\hbar}P_0\mathcal{F}_{xy}$, which produces the magnetization of Eq.~\eqref{M_z}.

It is worth emphasizing, that the obtained orbital magnetization and anomalous conductivity are extensive quantities, which at fixed plasma frequency $\gamma \sim g_0\sqrt{N}$ have finite values in the thermodynamics limit $N,S\rightarrow\infty$. This result is in stark contrast to the previously obtained thermodynamic observables of a gas of two-level systems in a cavity~\cite{pilar2020thermodynamics}, where it was shown that under the same condition all cavity induced corrections to the thermodynamic quantities vanish in the limit $N\to\infty$.

We have seen from the above macroscopic consideration that the ground state magnetization is produced by localized currents at the edges of a finite sample. Below we illustrate the appearance of these currents by considering a scattering of polaritonic states at the edge of a semi-infinite 2DEG that occupies a half plane $y\geqslant 0$. 

Let us first consider a one-electron plane wave polariton falling on the boundary at $y=0$. For the incident wave, $\Psi_{in}(\mathbf{r})=\Phi_{\mathbf{k}}(\mathbf{q})e^{i\mathbf{kr}}$, where $\Phi_\mathbf{k}(\mathbf{q})$ is defined by Eq.~\eqref{eq:Phi}, the photonic part is assumed to be in its ground state $n_{in}=l_{in}=0$. The reflected wave is the superposition of the elasically scattered wave with $n_{ref}=l_{ref}=0$ and inelastically scattered waves with $n_{ref},l_{ref}\neq 0$ which couple to the incident wave proportionally to $\rho^{2 n_{ref}+|l_{ref}|}$. Thus, to the leading order in $\rho$ the incident wave excites only the waves with $n_{ref}=0,l=-1,0,+1$. The coefficients in the linear superposition can be found from the condition that $\Psi|_{y=0}=o(\rho)$. The wavefunction for energy $E$ and wavevector $k_x$ can be then written as
\begin{align}
   & \Psi_{E,k_x}(y) =  \Phi_{0,k_x,-k_y}e^{-ik_yy} - \Phi_{0,k_x,k_y}e^{ik_yy} + \nonumber \\&ik_y\rho\left[ (1+\frac{\Delta}{\Omega_0})\Phi_{1,k_x,\kappa_{+}}e^{i\kappa_{+}y} - (1-\frac{\Delta}{\Omega_0})\Phi_{-1,k_x,\kappa_{-}}e^{i\kappa_{+}y}\right],
\end{align}
where the polaritonic factors $\Phi_{l,k_x,k_y}$ are given by Eq.~\eqref{eq:Phi} for $n=0$, $k_y =\sqrt{2m/\hbar^2E-k_x^2}$, $\kappa_{\pm}=\sqrt{2m/\hbar^2E-k_x^2-\epsilon_{0,\pm}}$, and $\epsilon_{n,l}$ are the corresponding harmonic oscillator energies.

From the wavefunction we can compute the charge current as  $j_x = \frac{e}{m}\langle \Psi | (\hbar k_x-\rho m\Omega q_x) |\Psi \rangle $. For the simplest estimate, we omit the photon-induced electron entanglement and treat the electrons as independent. The net current is then just the sum of all the currents produced by the electrons in the Fermi sea. In the case of sufficiently large Fermi energy $\epsilon_F$, when the paramagnetic contribution dominates, the current density reads
\begin{align}
    j_x(y) = \frac{e}{\hbar} \epsilon_{F} n \mathcal{F}_{xy}\frac{4 J_{3}(2k_F y)}{k_F y^2},
\end{align}
where $J_3(x)$ is the third order Bessel function of the first kind.
The edge current $I_{edge}=\int dy J(y)= \frac{e}{\hbar} \epsilon_{F} n \mathcal{F}_{xy}$  then agrees with the result obtained via the many-body calculation of magnetization using periodic boundary conditions if we recall that the nonintercating pressure $P_0\sim \epsilon_F n$. The expression for current density in the limit $\epsilon_F\ll \hbar\Omega_0$ can be rewritten as
\begin{align}
  &j(y)= nev_F\alpha \left(\frac{\epsilon_F}{\hbar\Omega}\right)^2\frac{\Delta}{\Omega_0} \mathrm{F}(k_Fy),~ \mathrm{F}(x)=\frac{J_3(2x)}{x^2},
\end{align}
where $\alpha$ is the fine structure constant.
As we can see the current oscillates and decays in the bulk of the structure as $(k_Fy)^{-5/2}$. 

We have shown that the coupling of electrons to the vacuum electromagnetic field of the gyrotropic cavity induces the macroscopic orbital magnetization of the electron gas, and correspondingly, leads to the emergence of the anomalous Hall conductivity and the existence of edge currents at the boundaries of the sample. In our analysis we have employed the translation invariance of the electronic subsystem. Apparently, the inevitable disorder breaking translation invariance may affect the magnetization. Moreover, local inhomogeneities in the electron gas would couple the centre of mass and the relative electron motion, and therefore the latter will also be influenced by the vacuum electromagnetic field. Quantitative estimation of the effect of disorder is an interesting problem for the future. Finally, it is worth mentioning that in a possible experiment, the gyrotropy in the mirrors is most naturally induced by using magnetic materials or applying an external magnetic field, which may generate the usual diamagnetic currents in the electron gas. While the full screening  of the electron system from the external and/or stray field is a challenging task, the effect induced by the vacuum fluctuations can be extracted by studying the dependence on the cavity photon frequency. 

To conclude, the extensive nature of the induced magnetization and the emergent edge currents indicates that the cavity engineering can be used to alter the macroscopic properties of the ground state of the low-dimensional electron systems, and thus further smears the boundaries between the fields of nanophotonics,  quantum optics and condensed matter theory.

\textit{Acknowledgements --}
We thank Ivan Sinev for the help with the figure preparation. The work of I.I. and D.G. was supported by The
was supported by the Russian Science Foundation
(project 20-12-00224).
I.V.T. acknowledges support by Grupos Consolidados UPV/EHU del Gobierno Vasco (Grant No. IT1249-19).

\bibliographystyle{plain}
\bibliography{references}

\newpage
\onecolumngrid
\setcounter{page}{0}
\setcounter{table}{0}
\setcounter{section}{0}
\setcounter{figure}{0}
\setcounter{equation}{0}
\renewcommand{\thepage}{\Roman{page}}
\renewcommand{\thesection}{S\arabic{section}}
\renewcommand{\thetable}{S\arabic{table}}
\renewcommand{\thefigure}{S\arabic{figure}}
\renewcommand{\theequation}{S\arabic{equation}}
\cleardoublepage
\vfill\eject
\thispagestyle{empty}

\section*{Supplemental Material}

\section{I. Derivation of the macroscopic magnetization}

The intrinsic magnetization can be determined by probing the system
with an external magnetic field ${\bf B}=\nabla\times{\bf A}$ and
analyzing the energy change as a function of ${\bf B}$ (in the following
we assume ${\bf B}=B\hat{\bm{z}}$). Specifically, we consider the Hamiltonian
\begin{equation}
H=\sum_{i=1}^{N}\frac{1}{2m}[\hat{{\bf p}}_{i}-g_0{\bf q}-\frac{e}{c}{\bf A}({\bf r}^{i})]^{2}+\frac{1}{2}\sum_{i\ne j}V_{{\bf r}^{i}-{\bf r}^{j}}+\frac{\hbar\Omega_0}{2}(\hat{\bm{\pi}}+\frac{\Delta}{\Omega_0}\hat{\bm{z}}\times{\bf q})^{2}+\frac{\hbar\Omega}{2}{\bf q}^{2},\label{eq:H-B}
\end{equation}
 where ${\bf A}({\bf r})$ corresponds to a uniform magnetic field
in $z$-direction and other notations are the same as in the main text. Then the ground state magnetization is computed
as follows (see, e.g. \cite{Shi2007})
\begin{equation}
M_{z}=-\frac{1}{S}\frac{\partial}{\partial B}{\rm Tr}[\rho(H-\mu\hat{N})]\Big|_{B=0}\label{eq:M-def1}
\end{equation}
where $\rho(\{{\bf r}_{1}^{i}\},\{{\bf r}_{2}^{i}\})$ is the ground
state N-body density matrix (here $\{{\bf r}^{i}\}\equiv{\bf r}^{1},{\bf r}^{2},...,{\bf r}^{N}$), and $\mu$ is the chemical potential. 

To compute the magnetization below we adopt and slightly generalize
the density matrix perturbation theory of Refs. \cite{Essin2010,GonZwa2011}.
First, we define the gauge (and translation) invariant N-body density
matrix
\begin{equation}
\tilde{\rho}(\{{\bf r}_{1}^{i}\},\{{\bf r}_{2}^{i}\})=\rho(\{{\bf r}_{1}^{i}\},\{{\bf r}_{2}^{i}\})e^{-i\frac{e}{c}\sum_{i=1}^{N}\int_{{\bf r}_{2}^{i}}^{{\bf r}_{1}^{i}}{\bf A}({\bf r})d{\bf l}}\label{eq:tilde-rho}
\end{equation}
The contour integrals in the exponent go, for each particle, along
straight lines from ${\bf r}_{2}^{i}$ to ${\bf r}_{1}^{i}$. These
integrals compensate the phase acquired by $\rho$ under the gauge
transformation of ${\bf A}$. They also make $\tilde{\rho}$ translation
invariant if $B$ is homogeneous.

The magnetization can be calculated from the first order correction
to the grand canonical potential
\begin{equation}
W_{1}={\rm Tr}[\delta\tilde{\rho}_{1}(H_{0}-\mu\hat{N})]=\sum_{n}(E_{n}-\mu N)\langle n|\delta\tilde{\rho}_{1}|n\rangle\label{eq:W1-def}
\end{equation}
 where $\delta\tilde{\rho}_{1}=\tilde{\rho}-\rho_{0}$ is the first
order in $B$ correction to the gauge invariant density matrix \ref{eq:tilde-rho},
and $H_{0}$, $|n\rangle$, $E_{n}$, and $\rho_{0}=|0\rangle\langle0|$
are, respectively, the Hamiltonian, its eigenstates, energies, and
the ground density matrix for $B=0$. 

The diagonal matrix elements $\langle n|\delta\tilde{\rho}_{1}|n\rangle$
can be found from the idempotency condition for the density matrix
\cite{Essin2010,GonZwa2011}. Namely, $\rho$ should satisfy the identity
$\rho\rho=\rho$. This translates to the following identity for $\tilde{\rho}$
\begin{equation}
\tilde{\rho}(\{{\bf r}_{1}^{i}\},\{{\bf r}_{2}^{i}\})=\int d\{{\bf r}_{3}^{i}\}\tilde{\rho}(\{{\bf r}_{1}^{i}\},\{{\bf r}_{3}^{i}\})\tilde{\rho}(\{{\bf r}_{3}^{i}\},\{{\bf r}_{2}^{i}\})e^{i\frac{e}{c}\sum_{i=1}^{N}\phi_{123}^{i}}\label{eq:idempotency1}
\end{equation}
where $\phi_{123}^{i}$ is a magnetic flux through a triangle formed
by the points ${\bf r}_{1}^{i}$, ${\bf r}_{2}^{i}$, and ${\bf r}_{3}^{i}$,
\begin{align}
\phi_{123}^{1} & =\int_{{\bf r}_{1}^{1}}^{{\bf r}_{2}^{i}}{\bf A}({\bf r})d{\bf l}+\int_{{\bf r}_{2}^{1}}^{{\bf r}_{3}^{i}}{\bf A}({\bf r})d{\bf l}+\int_{{\bf r}_{3}^{1}}^{{\bf r}_{1}^{i}}{\bf A}({\bf r})d{\bf l}=\int_{C_{123}^{i}}{\bf A}({\bf r})d{\bf l}\label{eq:flux}\\
 & ={\bf B}\cdot{\bf s}_{123}=-\frac{1}{2}{\bf B}\cdot[({\bf r}_{1}^{i}-{\bf r}_{3}^{i})\times({\bf r}_{3}^{i}-{\bf r}_{2}^{i})]
\end{align}
Using this expression and expanding Eq.\ref{eq:idempotency1} to the
first order in $B$ we get the following equation for $\delta\tilde{\rho}_{1}=\tilde{\rho}-\rho_{0}$
\begin{equation}
\delta\tilde{\rho}_{1}=\rho_{0}\delta\tilde{\rho}_{1}+\delta\tilde{\rho}_{1}\rho_{0}-i\frac{e}{2c}\text{\ensuremath{\varepsilon}}_{z\mu\nu}B\sum_{i=1}^{N}[r_{\mu}^{i},\rho_{0}][r_{\nu}^{i},\rho_{0}]\label{eq:delta-rho1}
\end{equation}
where we used the idempotency of the unperturbed density matrix, $\rho_{0}\rho_{0}=\rho_{0}$.
Notice that this equation is almost trivial N-body generalization
of the first order result for non-interacting particles \cite{Essin2010,GonZwa2011}.
Now we can find the required diagonal matrix elements,
\begin{align*}
\langle0|\delta\tilde{\rho}_{1}|0\rangle & =i\frac{e}{2c}\text{\ensuremath{\varepsilon}}_{z\mu\nu}B\sum_{i=1}^{N}\langle0|[r_{\mu}^{i},\rho_{0}][r_{\nu}^{i},\rho_{0}]|0\rangle,\qquad n=0,\\
\langle n|\delta\tilde{\rho}_{1}|n\rangle & =-i\frac{e}{2c}\text{\ensuremath{\varepsilon}}_{z\mu\nu}B\sum_{i=1}^{N}\langle n|[r_{\mu}^{i},\rho_{0}][r_{\nu}^{i},\rho_{0}]|n\rangle,\qquad n\ne0,
\end{align*}
 and the first order correction to the grand potential,
\begin{align}
W_{1} & =-i\frac{e}{2c}\text{\ensuremath{\varepsilon}}_{z\mu\nu}B\sum_{i=1}^{N}\Big\{\sum_{n\ne0}(E_{n}-\mu N)\langle n|[r_{\mu}^{i},\rho_{0}][r_{\nu}^{i},\rho_{0}]|n\rangle\nonumber \\
 & -(E_{0}-\mu N)\langle0|[r_{\mu}^{i},\rho_{0}][r_{\nu}^{i},\rho_{0}]|0\rangle\Big\}.\label{eq:W1-1}
\end{align}
Because in our case the center of mass (CM) and the relative motions
are independent it is convenient to separate the corresponding contributions
in the above equation,e
\begin{align}
W_{1} & =-i\frac{e}{2c}\text{\ensuremath{\varepsilon}}_{z\mu\nu}B\Big\{\sum_{n\ne0}(E_{n}-\mu N)\sum_{i}\langle n|[\xi_{\mu}^{i},\rho_{0}][\xi_{\nu}^{i},\rho_{0}]|n\rangle\nonumber \\
 & -(E_{0}-\mu N)\sum_{i}\langle0|[\xi_{\mu}^{i},\rho_{0}][\xi_{\nu}^{i},\rho_{0}]|0\rangle\Big\}\nonumber \\
 & -\frac{i}{2}\text{\ensuremath{\varepsilon}}_{z\mu\nu}B\Big\{\sum_{n\ne0}(E_{n}-\mu N)\langle n|[R_{\mu},\rho_{0}][R_{\nu},\rho_{0}]|n\rangle\nonumber \\
 & -(E_{0}-\mu N)\langle0|[R_{\mu},\rho_{0}][R_{\nu},\rho_{0}]|0\rangle\Big\},\label{eq:W1-2}
\end{align}
where ${\bf R}=\frac{1}{\sqrt{N}}\sum_{i=0}^{N}{\bf r}^{i}$, and
$\bm{\xi}^{i}={\bf r}^{i}-\frac{1}{\sqrt{N}}{\bf R}$. Separability of the CM assumes
that all eigenstates have a product form, $|n\rangle=|\chi_{r}^{{\rm rel}}\rangle|\Phi_{l}\rangle$
where $r$ and $l$ are the quantum numbers for the relative and the
CM motions, respectively. Obviously the ground state density matrix
also has a product form, $\rho_{0}=\rho_{0}^{{\rm rel}}\rho_{0}^{{\rm CM}}$,
where $\rho_{0}^{{\rm rel}}=|\chi_{0}^{{\rm rel}}\rangle\langle\chi_{0}^{{\rm rel}}|$
and $\rho_{0}^{{\rm CM}}=|\Phi_{0}\rangle\langle\Phi_{0}|$. Using
these properties we rewrite Eq.\ref{eq:W1-2} as follows,
\begin{align}
W_{1} & =-i\frac{e}{2c}\text{\ensuremath{\varepsilon}}_{z\mu\nu}B\Big\{\sum_{r\ne0}(E_{0}^{{\rm CM}}+E_{r}^{{\rm rel}}-\mu N)\sum_{i}\langle\chi_{r}^{{\rm rel}}|[\xi_{\mu}^{i},\rho_{0}^{{\rm rel}}][\xi_{\nu}^{i},\rho_{0}^{{\rm rel}}]|\chi_{r}^{{\rm rel}}\rangle\nonumber \\
 & -(E_{0}^{{\rm CM}}+E_{0}^{{\rm rel}}-\mu N)\sum_{i}\langle\chi_{0}^{{\rm rel}}|[\xi_{\mu}^{i},\rho_{0}^{{\rm rel}}][\xi_{\nu}^{i},\rho_{0}^{{\rm rel}}]|\chi_{0}^{{\rm rel}}\rangle\Big\}\nonumber \\
 & -i\frac{e}{2c}\text{\ensuremath{\varepsilon}}_{z\mu\nu}B\Big\{\sum_{l\ne0}(E_{l}^{{\rm CM}}+E_{0}^{{\rm rel}}-\mu N)\langle\Phi_{l}|[R_{\mu},\rho_{0}^{{\rm CM}}][R_{\nu},\rho_{0}^{{\rm CM}}]|\Phi_{l}\rangle\nonumber \\
 & -(E_{0}^{{\rm CM}}+E_{0}^{{\rm rel}}-\mu N)\langle\Phi_{0}|[R_{\mu},\rho_{0}^{{\rm CM}}][R_{\nu},\rho_{0}^{{\rm CM}}]|\Phi_{0}\rangle\Big\}\label{eq:W1-3}
\end{align}
Since the relative motion is time-reversal invariant, the first two
lines vanish and only the CM motion, which is coupled to the cavity
modes, can modify the energy at the linear order in $B$. Therefore
only the last two lines in \ref{eq:W1-3} contribute to the magnetization
of Eq.\ref{eq:M-def1},
\begin{align}
M_{z} & =-i\frac{e}{2cS}\text{\ensuremath{\varepsilon}}_{z\mu\nu}\Big\{\sum_{l\ne0}(E_{l}^{{\rm CM}}+E_{0}^{{\rm rel}}-\mu N)\langle\Phi_{l}|\partial_{P_{\mu}}\rho_{0}^{{\rm CM}}\partial_{P_{\nu}}\rho_{0}^{{\rm CM}}|\Phi_{l}\rangle\nonumber \\
 & -(E_{0}^{{\rm CM}}+E_{0}^{{\rm rel}}-\mu N)\langle\Phi_{0}|\partial_{P_{\mu}}\rho_{0}^{{\rm CM}}\partial_{P_{\nu}}\rho_{0}^{{\rm CM}}|\Phi_{0}\rangle\Big\}.\label{eq:M-1}
\end{align}
Here we use the momentum representation for the CM wave function and
perform the replacement $[R_{\nu},\rho_{0}^{{\rm CM}}]\to i\partial_{P_{\mu}}\rho_{0}^{{\rm CM}}\vert_{\mathbf{P=0}}$.
After the substitution $\rho_{0}^{{\rm CM}}=|\Phi_{0}\rangle\langle\Phi_{0}|$
and differentiation we get,
\begin{align*}
M_{z} & =-i\frac{e}{2cS}\text{\ensuremath{\varepsilon}}_{z\mu\nu}\Big\{\sum_{l\ne0}(E_{l}^{{\rm CM}}+E_{0}^{{\rm rel}}-\mu N)\langle\Phi_{l}|\partial_{P_{\mu}}\text{\ensuremath{\Phi}}{}_{0}\rangle\langle\partial_{P_{\nu}}\Phi_{0}|\Phi_{l}\rangle\\
 & -(E_{0}^{{\rm CM}}+E_{0}^{{\rm rel}}-\mu N)(\langle\partial_{P_{\mu}}\Phi_{0}|\partial_{P_{\nu}}\text{\ensuremath{\Phi}}{}_{0}\rangle-\langle\partial_{P_{\nu}}\Phi_{0}|\Phi_{0}\rangle\langle\Phi_{0}|\partial_{P_{\mu}}\text{\ensuremath{\Phi}}{}_{0}\rangle)\Big\}\\
= & i\frac{e}{2cS}\text{\ensuremath{\varepsilon}}_{z\mu\nu}\Big\{\sum_{l}(E_{l}^{{\rm CM}}+E_{0}^{{\rm rel}}-\mu N)\langle\partial_{P_{\mu}}\Phi_{0}|\Phi_{l}\rangle\langle\Phi_{l}|\partial_{P_{\nu}}\text{\ensuremath{\Phi}}{}_{0}\rangle\\
 & +(E_{0}^{{\rm CM}}+E_{0}^{{\rm rel}}-\mu N)\langle\partial_{P_{\mu}}\Phi_{0}|\partial_{P_{\nu}}\text{\ensuremath{\Phi}}{}_{0}\rangle\Big\}.
\end{align*}
last expression can be represented  more compactly
\begin{equation}
M_{z}=-\frac{e}{cS}{\rm Im}\langle\partial_{P_{x}}\Phi_{0}|\hat{H}_{0}^{{\rm CM}}+E_{0}^{{\rm CM}}+2E_{0}^{{\rm rel}}-2\mu N|\partial_{P_{y}}\text{\ensuremath{\Phi}}{}_{0}\rangle.\label{eq:M-final1}
\end{equation}
To proceed further to the thermodynamic limit we recall the solution
of the CM problem. The ${\bf P}$-dependent Hamiltonian, the ground
state energy, and the wave function are given by the following expressions:
\[
\hat{H}_{{\bf P}}^{{\rm CM}}=\frac{1}{2m}[\hat{{\bf P}}-g_0{\bf q}-\frac{e}{c}{\bf A}({\bf r}^{i})]^{2}+\frac{1}{2}\sum_{i\ne j}V_{{\bf r}^{i}-{\bf r}^{j}}+\frac{\hbar\Omega_0}{2}(\hat{\bm{\pi}}+\frac{\Delta}{\Omega_0}\hat{\bm{z}}\times{\bf q})^{2}+\frac{\hbar\Omega}{2}{\bf q}^{2},
\]
\[
E_{0}^{{\rm CM}}=\sqrt{\Omega^2+\Delta^{2}+\frac{N g_0^2}{m}},
\]
\[
\Phi_{0}({\bf P},{\bf q})=e^{-i\frac{\Delta}{\Omega_0}\rho[\mathbf{e}_z\times \mathbf{P}]\mathbf{q}}\psi_{n,l}(\mathbf{q}-\rho\mathbf{P}),~\rho= \frac{g_0\sqrt{N}}{m\Omega} = \sqrt{\frac{\hbar\Omega_0}{m}} \frac{\gamma}{\tilde{\Omega}_0^2+\gamma^2},~\gamma^2=\frac{g_0^2N\Omega_0}{\hbar m}.
\]
From here we may see that if $N g_0^2/m$ goes to finite value
in the thermodynamic limit, then $\hat{H}_{0}^{{\rm CM}}$, $E_{0}^{{\rm CM}}$,
and $N\langle\partial_{P_{x}}\Phi_{0}|\partial_{P_{y}}\text{\ensuremath{\Phi}}{}_{0}\rangle$
also have finite limits when $N,S\to\infty$. Therefore the first
two terms in Eq.\ref{eq:M-final1} go to zero in the thermodynamic
limit, and the macroscopic magnetization reduces to the following
compact form,
\begin{equation}
M_{z}=-\frac{2e}{c}P_{0}{\rm Im}\langle\partial_{P_{x}}\Phi_{0}|\partial_{P_{y}}\text{\ensuremath{\Phi}}{}_{0}\rangle=-2P_{0}\frac{e\Delta}{c\Omega_0}\rho^2\label{eq:M-final}
\end{equation}
 Here $P_{0}=(E_{0}^{{\rm rel}}-\mu N)/S$ is the ground state pressure
of the (interacting) 2D electron gas in the absence of photons.

\section{II. The case of Multimode cavity.}
In this section we consider the case of a multimode cavity, which supports arbitrary number of the modes, labeled by index $\alpha$ with frequency, coupling and gyration parameters $\Omega_{\alpha},g_{\alpha}$ and $\Delta_{\alpha}$, respectively. The centre of mass Hamiltonian then reads 
\begin{equation}
H_{{\bf k}}=\sum_{\alpha}\frac{\hbar\Omega_{\alpha}}{2}(\hat{\bm{\pi}}_{\alpha}+\Delta_{\alpha}/\Omega_{\alpha}\bm{e}_{z}\times{\bf q}_{\alpha})^{2}+\sum_{\alpha}\frac{\hbar}{2}\Omega_{\alpha}{\bf q}_{\alpha}^{2}+\frac{1}{2m}\Big(\hbar{\bf k}-\sqrt{N}\sum_{\alpha}g_{\alpha}{\bf q}_{\alpha}\Big)^{2}\label{eq:H_k-multimode}
\end{equation}
The last two terms can be rewritten as follows,
\begin{align}
\sum_{\alpha}\frac{\hbar}{2}\Omega_{\alpha}{\bf q}_{\alpha}^{2}+\frac{1}{2m}\Big(\hbar{\bf k}-\sqrt{N}\sum_{\alpha}g_{\alpha}{\bf q}_{\alpha}\Big)^{2}=
 \sum_{\alpha}\frac{\hbar\Omega_{\alpha}}{2}\left({\bf q}_{\alpha}-\tilde{g}_{\alpha}\hbar{\bf k}\right)^{2}+\frac{1}{2m}\Big(\sum_{\alpha}\sqrt{N}g_{\alpha}\left[{\bf q}_{\alpha}-\tilde{g}_{\alpha}\hbar{\bf k}\right]\Big)^{2}+\frac{\hbar^2{\bf k}^{2}}{2m^{*}}\label{eq:E_pot-multimode}
\end{align}
 where
\[
\tilde{g}_{\alpha}=\frac{\sqrt{N}g_{\alpha}}{m\hbar\Omega_{\alpha}}\qquad\text{and}\qquad m^{*}=m\left(1+\sum_{\alpha}\frac{Ng_{\alpha}^2}{\hbar m\Omega_{\alpha}}\right)
\]
This identity can be checked by the direct substitution.

The representation of Eq.(\ref{eq:E_pot-multimode}) implies the following:
if $\psi_{n}(\{{\bf q}_{\alpha}\})$ is an eigenfunction of the Hamiltonian
(\ref{eq:H_k-multimode}) with ${\bf k}=0$,
\[
H_{0}\psi_{n}(\{{\bf q}_{\alpha}\})=E_{n}\psi_{n}(\{{\bf q}_{\alpha}\}),
\]
then the function
\begin{equation}
\Phi_{n,{\bf k}}(\{{\bf q}_{\alpha}\})=e^{i\sum_{\alpha}\Delta_{\alpha}/\Omega_{\alpha}\rho_{\alpha}(\bm{e}_{z}\times{\bf q}_{\alpha}){\bf k}}\psi_{n}(\{{\bf q}_{\alpha}-\tilde{g}_{\alpha}\hbar{\bf k}\})\label{eq:Phi_k-multimode}
\end{equation}
is an eigenfunction of the full ${\bf k}$-dependent Hamiltonian
\[
H_{{\bf k}}\Phi_{n,{\bf k}}(\{{\bf q}_{\alpha}\})=\Big(E_{n}+\frac{{\hbar^2\bf k}^{2}}{2m^{*}}\Big)\Phi_{n,{\bf k}}(\{{\bf q}_{\alpha}\})
\]

Then the ground state Berry curvature for the multimode case is obtained
practically immediately 
\begin{equation}
\mathcal{F}_{xy}=2\sum_{\alpha}\frac{\Delta_{\alpha}}{\Omega_{\alpha}}\rho_{\alpha}^2.
\end{equation}

\end{document}